\documentclass[preprint,12pt]{elsarticle} 

\usepackage{amssymb}
\usepackage{amsmath}
\usepackage{caption}
\usepackage{subcaption}
\usepackage{graphicx}
\usepackage{xcolor}
\usepackage{booktabs}
\usepackage{subcaption}

\usepackage[section]{placeins}

\usepackage{mathtools}
\usepackage[normalem]{ulem}

\journal{NIM Section A}

\begin{document}

\begin{frontmatter}



\title{Improving Peak-Based Nuclide Identification in HPGe $\gamma$-Spectrometry
with Machine Learning and SHAP} %% Article title


\author{Samuel Emmons\corref{cor1}}
\cortext[cor1]{samuel.emmons@pnnl.gov}
\author{Kelly Truax}
\author{Maurice Lonsway}
\author{Bruce Pierson}
\author{Brian Archambault}

\affiliation{organization={Pacific Northwest National Laboratory},
	%Department and Organization
            %addressline={}, 
            city={Richland},
            postcode={99354}, 
            state={WA},
            country={USA}
            }

%% Abstract
\begin{abstract}
High-purity germanium gamma spectra often 
require time-consuming analyses from subject matter experts. 
Photopeaks within these spectra are carefully fitted and 
numerical methods are employed to assist with nuclide identification (NID) and
quantification. Amending the list of nuclides identified by analysis software 
can be nontrivial. When many samples 
need to be analyzed, it is therefore challenging to make timely \textit{and} correct 
decisions. Supervised machine-learning-based NID can serve as an 
expert-informed, automated tool to improve the initial set of radionuclides 
suggested to an analyst and more effectively drive subsequent quantification. 
To that end, we implemented machine learning
models that map photopeaks carefully fitted by analysts to NID results for 
experimental spectra
containing various isotopic combinations drawn from a set of 
65 isotopes. The best model achieved an F1 score of $0.97$, markedly surpassing the 
F1 score of 0.84 achieved by traditional software when compared using a nuclide 
library comprising the same 65 isotopes assessed by the models. 
Finally, we illustrated the most important input features for model predictions using
Shapley Additive Explanations. These explanations revealed that
the models use physically relevant photopeaks when making predictions for 
the isotopes in our nuclide library. 
\end{abstract}


\begin{highlights}
\item XGBoost and DNNs outperform traditional methods for NID. 
%and can enhance analyst workflow.
\item XGBoost models give higher F1 scores than DNNs for peak-based NID. 
\item ML models employed physically relevant photopeaks as shown by SHAP.
\end{highlights}

%% Keywords

\end{frontmatter}

\newpage
\section{Introduction}
\label{sec:Intro}
High-purity germanium (HPGe) detectors are integral to 
the assay of complex radioactive samples. The energy resolution of these detectors
enables the simultaneous observation of many distinct photopeaks in gamma spectra. 
For an analyst, identifying and quantifying radionuclide contributions to those spectra  
presents challenges typically overcome through years of training and help from the best
available commercial software. Still, this is a time-intensive process, and 
improvements to automated analyses will help spectroscopists quickly
\textit{and} correctly characterize complex samples.

The photopeaks in a spectrum can be automatically located using techniques such as
that of Mariscotti \cite{Mariscotti1967}, though analyst review and refitting is generally
required in complicated spectra with overlapping peaks. The counts  
in these peaks, together with information such as the
spectrum live time and the photopeak efficiency function for the measurement, form 
the basis for most contemporary nuclide identification (NID) and activity quantification
methods that are based on the work of Gunnink and Niday \cite{Gunnink1972} and 
Koskelo et al.~\cite{Koskelo1981}. 

The set of photopeaks may be considered as a physics-informed dimensional 
reduction of spectral information that is useful as an input to an NID
algorithm, including machine learning (ML) methods as in Ref.~\cite{Yoshida2002}.
A number of other techniques have also been used to reduce 
the dimensionality of gamma spectra prior to their use in ML algorithms carrying out
various regression and classification
tasks \cite{Pilato1999,Hague2019,Kamuda2019,Bandstra2023,Khatiwada2023,Lonsway2026}.
Some information is lost in such a reduction, but this approach has the advantage
of requiring fewer model parameters than methods making use of complete spectra as
inputs and consequently requires less data for training. 

Supervised ML methods that map spectral features to outputs (e.g., 
isotope labels or proportions) may have an advantage over non-ML numerical 
methods since they learn from previous expert decisions to complete the 
best possible mapping. Practically, these expert decisions are made when large 
radionuclide libraries are used in the NID process and an analyst must  
down-select the isotopes present in a complex sample (or sometimes add in 
isotopes missed or rejected by even the best commercial or custom software).

A perhaps more difficult, but still tractable, problem consists of using
complete spectra as inputs to carry out \textit{multi-label proportion prediction}. In this 
regression problem, the proportional contribution of each isotope in a list of 
possible isotopes to an observed spectrum is computed. Thus, classification 
and quantification are effectively carried out simultaneously, as in 
Refs.~\cite{Pilato1999,kamuda2020,Omen2024,Kim2025,Phan2026}. 
Challenges arise when, for example, one or more isotopes contribute at 
a proportion multiple orders of magnitude below the strongest contributors or when the 
level of source attenuation is not well characterized.

Regardless of the input type one uses when training ML models to map spectral
data to some desired target(s), one must have some 
body of accurately characterized data for the models to learn from. 
In that vein, many works have 
relied upon simulations \cite{Bandstra2023,Phan2026,Kamuda2017,Bilton2021}, 
though some have made use of both simulated \textit{and} experimental measurements 
\cite{Khatiwada2023,Omen2024,Kim2025,Daniel2020,Barradas2025,Lalor2026}. 

In this work, we used a set of $\sim$1600 well-labeled \textit{experimental} 
HPGe gamma spectra collected in various measurement geometries on tens of instruments 
of different types (e.g., coaxial, semi-planar, planar) at Pacific Northwest 
National Laboratory (PNNL). The photopeaks in these 
spectra were located 
and fitted using commercial spectroscopic software and then reviewed and 
refitted by expert analysts. We then applied neural networks and extreme 
gradient-boosted decision trees (XGBoost, or XGB) \cite{Chen2016} to the task of mapping 
photopeak areas from HPGe gamma spectra to NID results for 65 radioisotopes,
for which there is precedent in
Refs.~\cite{Khatiwada2023,Daniel2020,Galib2021,Romo2021,Sun2025}.

We selected XGB since it is apt for tabular data \cite{Shwartz2022} (such as 
a list of photopeak areas extracted from a spectrum). We compared its performance 
to that of dense neural networks (DNNs), as 
DNNs' propensity for learning patterns and correlations across data made them good
candidates for this multi-label scenario \cite{Huang2013,Chen2022,Tarekegn2024}. 
We then generated ML prediction explanations using SHAP (SHapley Addative 
exPlanations)~\cite{Lundberg2017,Lundberg2020}. Such methods have been 
applied to ML models mapping complete spectra to desired outputs, but we believe this is 
the first publication in which these tools are used to explain  
the importance of specific photopeaks in ML predictions. In an additional important 
step that is seldom taken in contemporary ML work, we compare the model NID capabilities 
to those of commercial software.

We describe the data used in this study in Sec.~\ref{sec:Data}. The ML models  
and a description of model parameter optimization are presented in Sec.~\ref{sec:Models}.
Then, nuclide prediction results and the demonstrated reliance of those predictions on 
relevant spectral photopeaks are presented in Sec.~\ref{sec:Results}. There, we also 
compare ML model NID performance to that of commercial spectroscopic software. 
Finally, we describe key conclusions and possible directions of future research 
in Sec.~\ref{sec:Sum}.

\section{HPGe Data}
\label{sec:Data}
The machine learning models doing multi-label classification in this study 
were developed and tested using $\sim$1600 
gamma spectra collected experimentally at PNNL during nuclear forensics (NF)
research and development (R\&D) activities. In those activities, radiochemical 
separations isolate radioisotopes of certain elements for 
quantification by gamma spectrometry. Consequently, a variety of data has 
been generated, and there are more than 800 unique combinations of isotopes in our 
spectra. Additionally, a holdout set of 123 spectra obtained
from recent NF R\&D work was employed to test the generalizability of our trained
models. 

The \textit{support} for each isotope is not uniform in the dataset. This
relative frequency of occurrence in the dataset ranged from about $1\%$ to $48\%$. Further, spectra with a handful of isotopes were more prevalent in the 
dataset than those with many (a typical spectrum had $\sim 9$ isotopes
present).

The data was first analyzed within the Apex-Gamma$^{\text{TM}}$
Lab Productivity Suite~\cite{Canberra2017}. 
Peaks were carefully fitted with the Interactive Peak
Fit software prior to NID. Analysts often down-selected from, but
occasionally added to, a list of radionuclide identifications produced within 
the software for the various spectra. Photopeak energies and areas (counts within 
each peak), the NID list, and other metadata and quantification 
information were organized in a data structure for this study. 

\begin{figure}[t]
	\centering
	\includegraphics[width=9cm]{"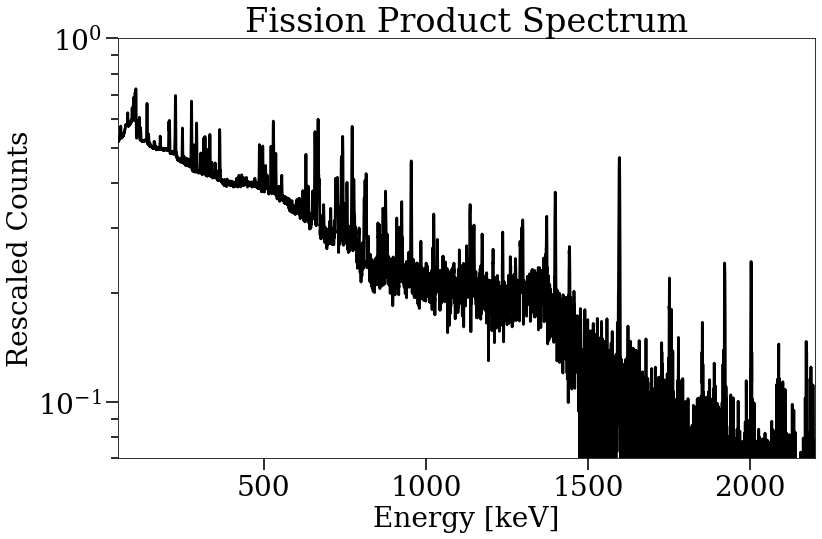"}
	\caption{Gamma spectrum from which peaks were extracted for
		model input comparison.}
	\label{fig:Spectrum1}
\end{figure}

To prepare the photopeaks for use as ML inputs, the set of photopeak areas for each 
spectrum was mapped onto an input vector with 359 entries corresponding to a fixed set 
of known gamma emission energies of the 65 isotopes in our NID list. This was done as 
follows:
\begin{enumerate}
	\item The elements of each input vector were initialized to zero. Each entry was
	labeled with a relevant photopeak energy (if two important photopeak energies were
	within 0.5~keV of each other, a single bin was formed and labeled with the average
	of the two energies). 
	\item Each photopeak area in a spectrum was added to the vector entry that its peak 
	centroid value was nearest (within a tolerance of 0.7 keV below 600~keV or 1.0~keV at
	or above 600~keV). For example, a peak with centroid 311.7~keV would be placed into
	the bin corresponding to the energy 311.9~keV for Pa-233 in our input space.
	\item The photopeak areas in each input vector were re-scaled such that the new 
	vector of values resided on $(0,1)$ and had values spanning fewer orders of magnitude 
	than the original range of areas.
\end{enumerate}
In the logarithmic rescaling method we used, the $i^{\text{th}}$ photopeak area $s_{i}$ becomes
\begin{equation}
	\label{eq:rescale1}
	\hat{s}_{i}=-\frac{s^{\prime}_{i}}{\text{min}(\textbf{s}^{\prime})}+1\,,
\end{equation}
where 
\begin{equation}
	\label{eq:rescale2}
	s^{\prime}_{i}=\log_{10}\left( \frac{s_{i}+1}{\Sigma_{j}(s_{j}+1)}\right)\,.
\end{equation}
This was described previously in Ref.~\cite{Chaouai2022}, where it was applied to 
complete spectra. The output of Eq.~\eqref{eq:rescale2} is negative and is 
therefore scaled and shifted in Eq.~\eqref{eq:rescale1}. We found that scaling in
this fashion enabled models to learn more quickly and converge to better
solutions (i.e., lower loss on validation and test data) than models trained on 
photopeak areas that were either directly min-max scaled to $[0,1]$ \textit{or} 
divided by the photopeak efficiency at their respective energies 
and then min-max scaled to [0,1].

\begin{figure}[h]
	\centering
	\includegraphics[width=9cm]{"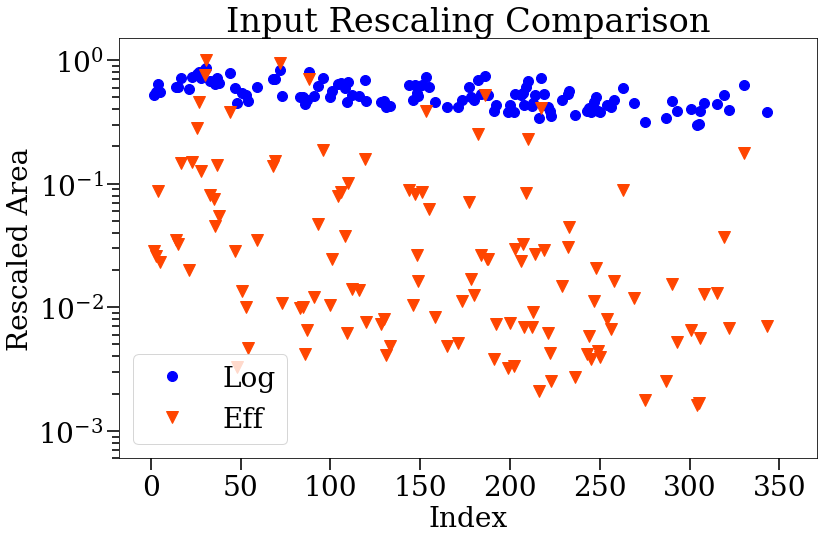"}
	\caption{Comparison of logarithmic (see Eqs.\eqref{eq:rescale1} and 
		\eqref{eq:rescale2}) and area-by-efficiency rescaling of peak
		data plotted versus the peak index in the model input vector.}
	\label{fig:Rescale}
\end{figure}
 
An example spectrum containing photopeaks from many fission products is shown
in Fig.~\ref{fig:Spectrum1}, and a comparison of its log-rescaled peaks to the 
peaks rescaled by dividing first by photopeak efficiency at
each energy and then min-max scaling the values onto $[0,1]$ is shown in 
Fig.~\ref{fig:Rescale}. Even when scaled by 
peak efficiency, the photopeak areas (red-orange triangles) span nearly three orders 
of magnitude. Rescaling by Eqs.~\eqref{eq:rescale1} and \eqref{eq:rescale2} condenses the inputs into a narrower band (blue dots) and improves computational stability while preserving general relationships between peak sizes.

\section{Model Architectures and Optimization}
\label{sec:Models}

We used three architectures to predict NID: a multi-label DNN, an ensemble of 
binary relevance (BR) DNNs (one binary classifier per target nuclide), and XGB
classifiers.
To account for the varied nuclide support while training the models, 
we performed a multi-label stratified split that maintains the relative label distribution
across the training, validation, and test sets. Mean model performance metrics 
were computed with 5-folds cross validation, since results can differ from one split to the
next. 80\% of the data was used for training/validation and the other 20\% was withheld as
a test set. 

\begin{table}[h]
	\centering
	\begin{tabular}{|l|}
		\hline
		\textbf{XGBoost}                                                                                    \\ \hline
		\begin{tabular}[c]{@{}l@{}}N estimators: [100, 200]\\ Max depth: 3, 6, 10\\ 
			Learning rate: [0.05, 0.3]\\ Subsample ratio: [0.7, 1.0] 
		\end{tabular}      
		\\ \hline
		
		\textbf{DNN}                                                                                                                                                             \\ \hline
		\begin{tabular}[c]{@{}l@{}}Activation:  `relu', `selu', `elu', `tanh'\\ 
			Optimizer:  `adam', `adagrad', `adamax', `nadam'\\ Hidden layers: [1, 5]\\
			Neurons/layer: [32, 256]\\ Dropout: [0.0, 0.2]\\ Learning rate: [1e-3, 1e-2]\\
			batch size: [16, 128]\end{tabular} \\ \hline
	\end{tabular}
	\caption{Hyperparameter spaces searched for the XGB and DNN models.}
	\label{tab:hyperparams}
\end{table}

\subsection{Model Hyperparameters}
\label{subsec:hyper}
Each model has structural parameters that were set \textit{prior to training}. 
For example, the number of estimators, maximum tree
depth, the learning rate, and the ratio of data selected (subsampled) to data held out 
in the training of each successive tree may be adjusted in an XGB model. In a DNN, 
the nonlinear activation function used between hidden layers, the optimizer used
to adjust model weights during training, the number of hidden layers, the number of 
neurons per layer, and other parameters may be adjusted to 
improve a model's fitness for the mapping task at hand (i.e., NID). These 
are known as \textit{hyperparameters}, and it is not possible to predict the best
combination of them \textit{a priori}. Therefore it is necessary to explore 
various combinations, train each resulting model's \textit{internal}
parameters, and measure model performance against a common metric.

To obtain optimal model architectures, we used the Bayesian 
optimization methods of Ref.~\cite{Nogueira2014}. We also tested a 
random search, a grid search, and Optuna's \cite{Akiba2019} implementation 
of Tree-structured Parzen Estimators (also a Bayesian method) but did not observe 
them outperform the implementation of Ref.~\cite{Nogueira2014}. 

When optimizing the ML models, 50 initial points for the Bayesian algorithm
were selected uniform-randomly from the hyperparameter space defined by the lists and 
intervals shown in Tab.~\ref{tab:hyperparams}. Then, 50 more optimization iterations 
were completed to minimize the binary cross entropy (BCE) objective function (i.e., 
find optimal hyperparameters).  
To prevent over-fitting for each model, training ended when model performance 
stopped improving on a validation set for seven training iterations using the 
\textit{early stopping} method.

\begin{table}[h]
	\centering
	\begin{tabular}{@{}lllllllll@{}}
		\toprule
		
		\textbf{}        & \textbf{Recall} & \textbf{Precision} & \textbf{F1}  \\ 
		\textbf{XGB} & 0.956(6)       & 0.978(6)         & 0.967(6)      \\ 
		\textbf{DNN$_{\text{BR}}$}         & 0.936(4)       & 0.969(2)     & 0.952(3) \\
		\textbf{DNN}                       & 0.938(4)       & 0.957(7)     & 0.947(5) \\
		
		\textbf{Genie 2k} & 0.955          & 0.748            & 0.839     \\

		\bottomrule
	\end{tabular}
	\caption{Recall, precision, and F1 score listed for each of three different model approaches: multi-label DNN, binary relevance DNNs, and binary relevance XGB classifiers (plus or minus uncertainty in parentheses). The final row lists the 
	metrics for the NID predictions made by Genie 2000 on the complete dataset.}
	\label{tab:metrics}
\end{table}

\section{NID Predictions and Explanations}
\label{sec:Results}
Predictions made by the optimal models produced in the hyperparameter search of 
Sec.~\ref{subsec:hyper} were assessed using the F1 score, i.e., the harmonic mean
of prediction precision (ratio of true positives (TP) to TP \textit{and} 
false positives (FP)) and recall (ratio of TP to TP
\textit{and} false negatives (FN)). Precision
and recall were computed using total number of TP, FP, and FN across the dataset.
We also compared ML model predictions for the holdout data 
mentioned in Sec.~\ref{sec:Models} to the NID predictions made automatically by 
Genie 2000 using a nuclide library consisting of the same 65 radionuclides assessed by 
the ML models. The peak-matching tolerance was set to 1.0~keV.

\begin{figure}[h]
	\centering
	\includegraphics[width=9cm]{"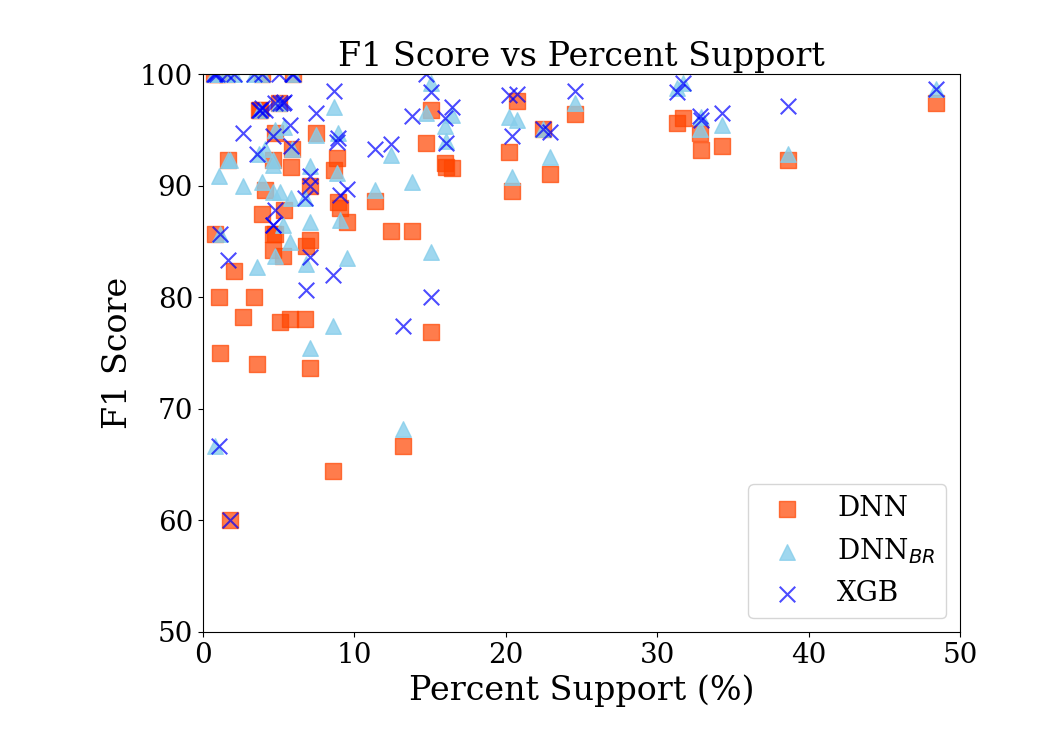"}
	\caption{F1 score versus percent support within the dataset for each nuclide.} 
	\label{fig:fnfp_PosPer}
\end{figure}

\subsection{Model Prediction Overview}
The recall, precision, and F1 score evaluated on 
the test data are listed in Tab.~\ref{tab:metrics} for the XGB classifiers, the 
DNN$_{\text{BR}}$ classifiers, the multi-label DNN, and Genie 2000. 
The F1 scores are similar, but XGB was generally the best 
model regardless of variations due to model 
parameter seeding or dataset shuffling differences. The uncertainties in
the table are the standard deviation on the mean computed in the 5-folds cross
validation process. 

The ML models outperformed the NID methods of Genie 2000 primarily by producing far 
fewer false positives. When making predictions for the holdout set of 123 spectra,
the best model achieved an overall $\text{F}1=0.92$, while G2k achieved an
$\text{F}1=0.80$. Central to this achievement was the fact that the models relied upon 
not only the correct photopeaks for a prediction, but also the context(s) in which 
those photopeaks typically occur.

For each model, misclassifications were more likely when
the support for a given nuclide was low. In Fig.~\ref{fig:fnfp_PosPer}, the F1 score 
for each nuclide label is plotted versus the percent support in the data for the 
multi-label DNN, DNN$_{\text{BR}}$, and XGB models. Generally, with 
reduced support there is a reduction in the F1 score for all models.

\begin{table}[t]
	\centering
	\begin{tabular}{@{}llll|lc@{}}
		\toprule
		
		\textbf{} & \textbf{Recall} & \textbf{Precision} & \textbf{F1} & \textbf{Isotope} & \textbf{Support} \\ 
		\hline
		
		\textbf{XGB}     & 0.84       & 1.00       & 0.91    & Cd-109 & 9\% \\
		\textbf{DNN$_{\text{BR}}$}    & 0.74      & 1.00      & 0.85   &   \\
		\textbf{DNN}      & 0.84       & 1.00      & 0.91  \\
		\textbf{Genie 2k} & 1.00       & 0.45         & 0.62 &		     \\
		
		\hline
		\textbf{XGB} & 0.92 & 1.00 & 0.96 &	 Nb-97 & 12\% \\ 
		
		\textbf{DNN$_{\text{BR}}$} & 0.92 & 1.00 & 0.96 & \\
		\textbf{DNN} & 1.00 & 1.00 & 1.00  \\
		 \textbf{Genie 2k} & 0.92 & 0.71 & 0.80 & \\
		
		\hline 
		\textbf{XGB} & 0.75 & 0.75 & 0.75 &	 Y-93 & 3\% \\
		\textbf{DNN$_{\text{BR}}$} & 1.00 & 0.80 & 0.89 & \\
		\textbf{DNN} & 0.75 & 0.75 & 0.75 & \\
		\textbf{Genie 2k} & 1.00 & 0.57 & 0.73 & \\
		
		\hline 
		\textbf{XGB} & 0.92 & 0.92 & 0.92 &	 Overall & \\
		\textbf{DNN$_{\text{BR}}$} & 0.87 & 0.91 & 0.89 & &  \\
		\textbf{DNN} & 0.87 & 0.89 & 0.88 & & \\
		\textbf{Genie 2k} & 0.97 & 0.68 & 0.80 &   & \\

		\bottomrule
	\end{tabular}
	\caption{Comparison of recall, precision, and F1 scores for model predictions for 
		Cd-109, Nb-97, and Y-93 in the holdout dataset to the same metrics for Genie 2000
		predictions. }
	\label{tab:SpecificF1s}
\end{table}

\begin{figure}[h]
	\centering
	\includegraphics[width=9cm]{"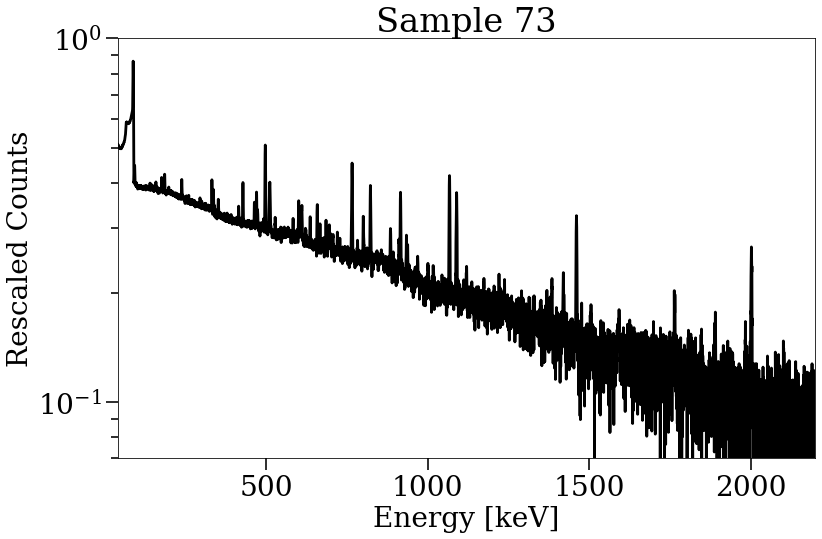"}
	\caption{Rescaled spectrum containing Cd-109 (88-keV photopeak) and other isotopes.}
	\label{fig:Cd109_spec}
\end{figure}

\subsection{Specific Isotope Predictions}
\label{sec:Samples}
To further demonstrate the efficacy the ML models used in this work,
we present a comparison of their nuclide predictions with those of 
Genie 2000 for isotopes representative of the following three scenarios: 
\begin{enumerate}
	\item Isotope with single likely gamma ray emission that has an energy close to
	a gamma- or X-ray of another isotope.
	\item Short-lived daughter radionuclide whose parent must usually be present. 
	\item Isotope rarely occurred in the training and validation data.
\end{enumerate}
The isotopes selected to illustrate 
these categories are Cd-109, Nb-97, and Y-93, respectively. In each case, we found
that XGB and DNNs 
tend to outperform the NID approach implemented in Genie 2000 based on the 
methods of Refs.~\cite{Gunnink1972,Koskelo1981}. This is because 
ML models learn to account for important spectroscopic context(s) 
that template-based methods do not, as 
shown in the SHAP explanations of Sec.~\ref{sec:Explain} below. We note that
the false negatives for Cd-109 arose for samples in which Cd-109 existed
in the presence of higher activity concentrations of isotopes of Ag. Fewer than 10 out of 1600 
samples in the development data had such a combination, and we expect this is why the 
predictions fell short of the threshold for a few similar samples in the 
holdout spectra. Otherwise, the models successfully distinguished when a
photopeak near 88~keV belonged to Cd-109 or, for example, a Pb X-ray.

A comparison of Recall, Precision, and F1 scores is shown for the selected isotopes 
in Tab.~\ref{tab:SpecificF1s}. 
Even when data support was low (as for Y-93 with 2.6\,\% support), ML models 
generally produced better NID F1 scores than Genie 2000. As described previously, the 
comparison was made fair by setting a nuclide 
identification threshold of 1.0 keV and using a library of the same 65 photopeaks 
with the same photopeak energies the ML models learned from. Table~\ref{tab:SpecificF1s}
also lists metrics for Genie 2k and the three ML models computed using the total number of
TP, FP, and FN across the complete holdout dataset (as opposed to weighted mean metrics).

\subsection{Feature-based Prediction Explanations}
\label{sec:Explain}
	
For the ML models of this work to be trusted with making isotope predictions, 
they must demonstrably depend upon physically relevant photopeaks within each spectrum. 
We used SHAP (SHapley Additive exPlanations) 
to show that our XGB and DNN models  rely on photopeaks associated
not only with the identified isotope, but also those corresponding to other aspects of
the spectral context (e.g., presence of a parent isotope peak lends credence to 
daughter nuclide prediction). 

\begin{figure}[t]
	\centering
	\includegraphics[width=9cm]{"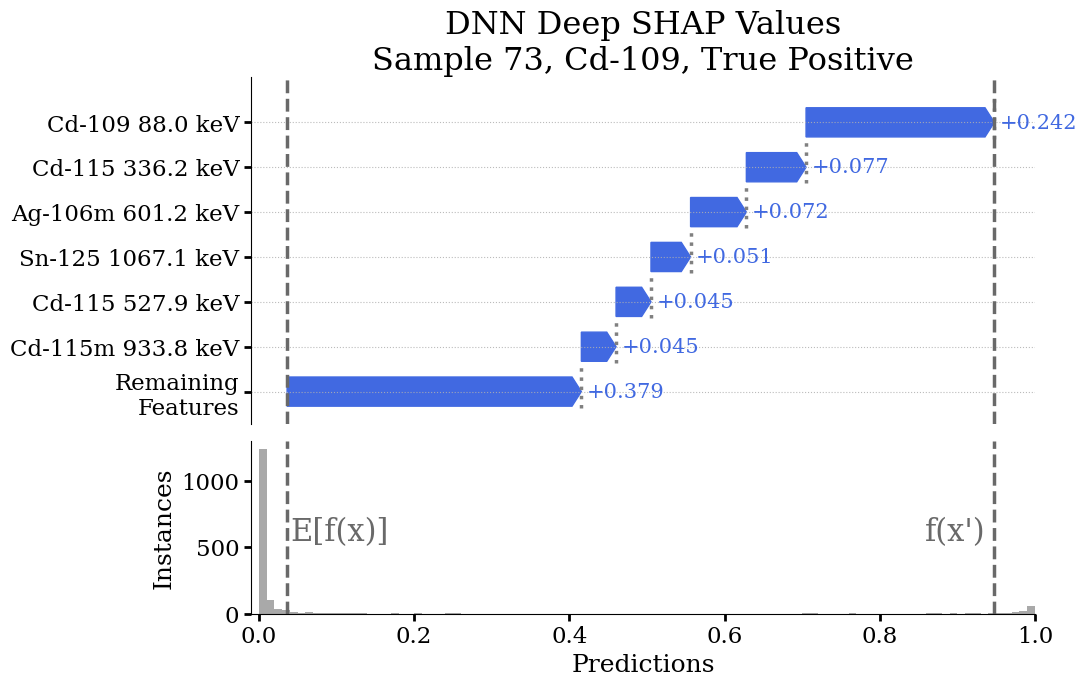"}
	\caption{Deep SHAP values for Cd-109 prediction made by multi-label DNN.}
	\label{fig:Cd109_Deep}
\end{figure}

\begin{figure}[h]
	\centering
	\includegraphics[width=9cm]{"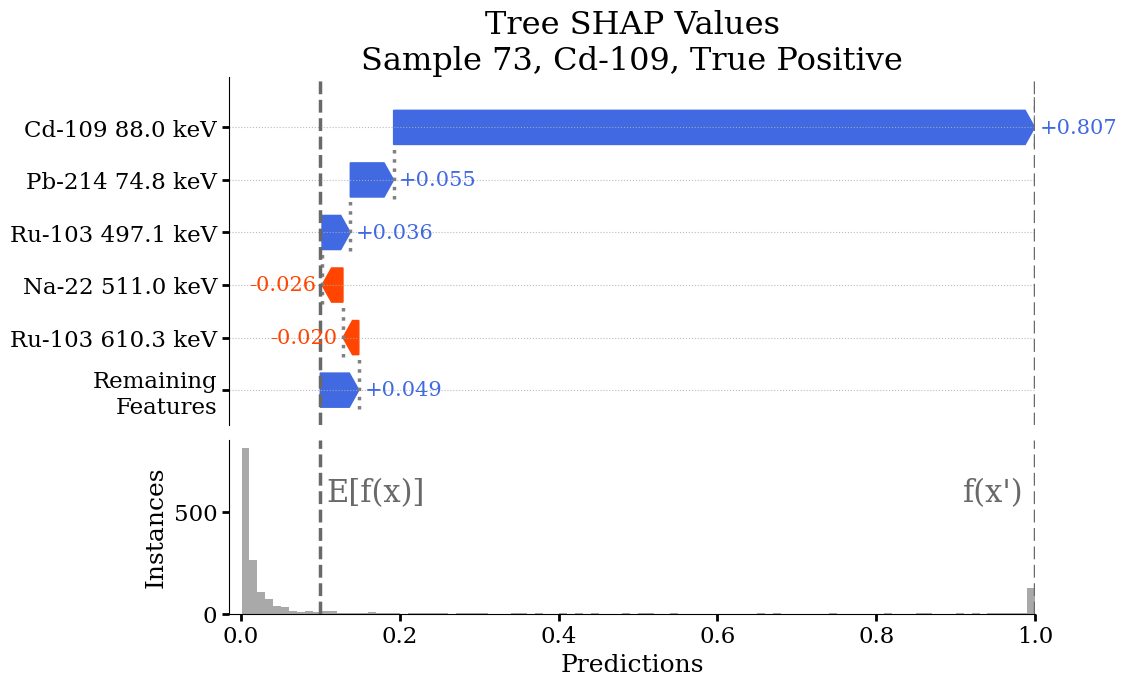"}
	\caption{Tree SHAP values for Cd-109 prediction made by XGB model.}
	\label{fig:Cd109_SHAP}
\end{figure}

SHAP assigns importance to features for predictions made by our multi-label and 
binary-relevance classifiers. To do so, SHAP uses a trained model and selected
\textit{background} dataset to compute expected values of nuclide predictions. Next, 
the user supplies a sample to evaluate. For an input $x$, the sum of feature 
SHAP values is added to the expected prediction value to obtain again the original
prediction value, $f(x)$. That is,
\begin{equation}
	\label{eq:SHAP}
	f(x) = \phi_{0}(f)+\sum_{i=1}^{M}\phi_{i}(f,x)\,,
\end{equation}
where the sum iterates over the $M$ input features and $\phi_{0}$ is the prediction 
\textit{expected value} for the isotope across the background dataset. For DNNs, we 
computed the feature contributions
$\phi_{i}$ using Deep SHAP as described in Ref.~\cite{Lundberg2017}, and for XGB models, we 
computed them using Tree SHAP with interventional feature perturbation \cite{Lundberg2020}.

\begin{figure}[h]
	\centering
	\includegraphics[width=9cm]{"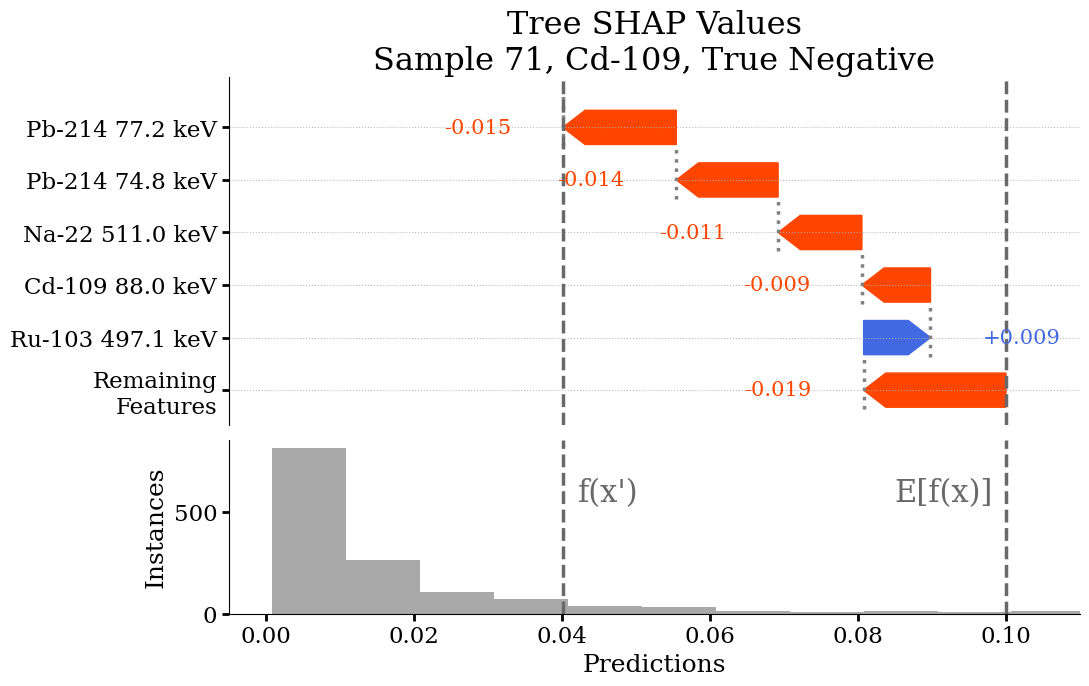"}
	\caption{Tree SHAP values when no Cd-109 is present ($\text{pred}=0$). }
	\label{fig:noCd109_SHAP}
\end{figure}

In this section, SHAP explanations of Cd-109 and Nb-97 model predictions are given.
Those for Y-93 are shown in Sec.~\ref{sec:limitations}. 
Additional SHAP explanations for three isotopes of Ce that commonly occur 
\textit{together} in spectra are presented in \ref{sec:AppendixA}.
Figures~\ref{fig:Cd109_Deep} and~\ref{fig:Cd109_SHAP} show SHAP 
explanations of multi-label DNN and XGB model predictions of Cd-109 for the spectrum 
shown in Fig.~\ref{fig:Cd109_spec}. In each figure, 
the lower panel is a histogram of prediction 
probabilities produced by each model for the background dataset used in the generation
of the SHAP values. The concentration of values near 0.0 shows that most of the time, 
Cd-109 was not predicted. The dashed line labeled $\text{E}[f(x)]$ is the expected 
value of the model prediction for Cd-109, and the SHAP values (horizontal directed bars
in the plots) describe the importance of each feature in
driving the specific sample prediction away from the expected value towards 0 or 1.

The models differed in the way they leveraged input features for  
Cd-109 predictions. For instance, the collection of `Remaining
Features' was more important for the multi-label DNN prediction than for the 
XGB prediction. Further, the XGB model learned to identify Cd-109 primarily based on 
the 88~keV peak size relative to nearby background peaks (see true negative prediction of 
Fig.~\ref{fig:noCd109_SHAP}, for example), while the DNN learned to identify it in relation 
to chemically similar isotopes in this example.

\begin{figure}[t]
	\centering
	\includegraphics[width=9cm]{"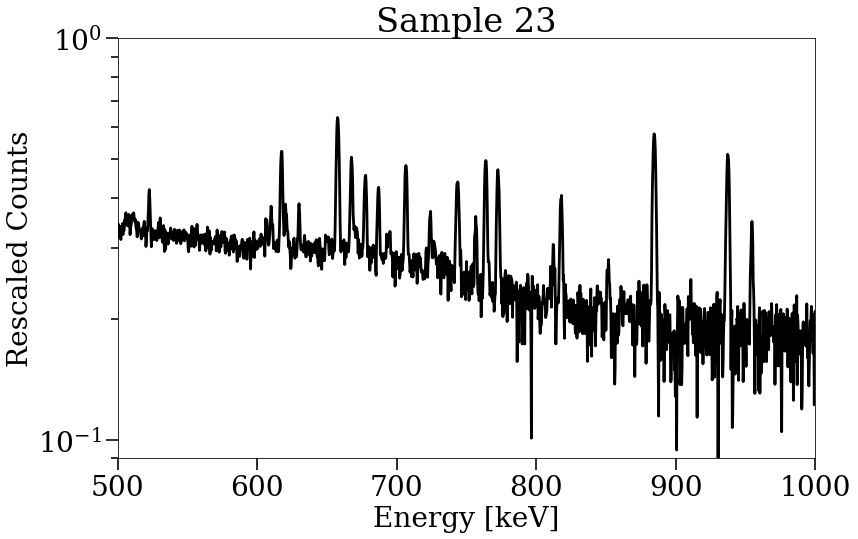"}
	\caption{Rescaled spectrum containing Nb-97 and other isotopes.}
	\label{fig:Nb97_spec}
\end{figure}

\begin{figure}[h]
	\centering
	\includegraphics[width=9cm]{"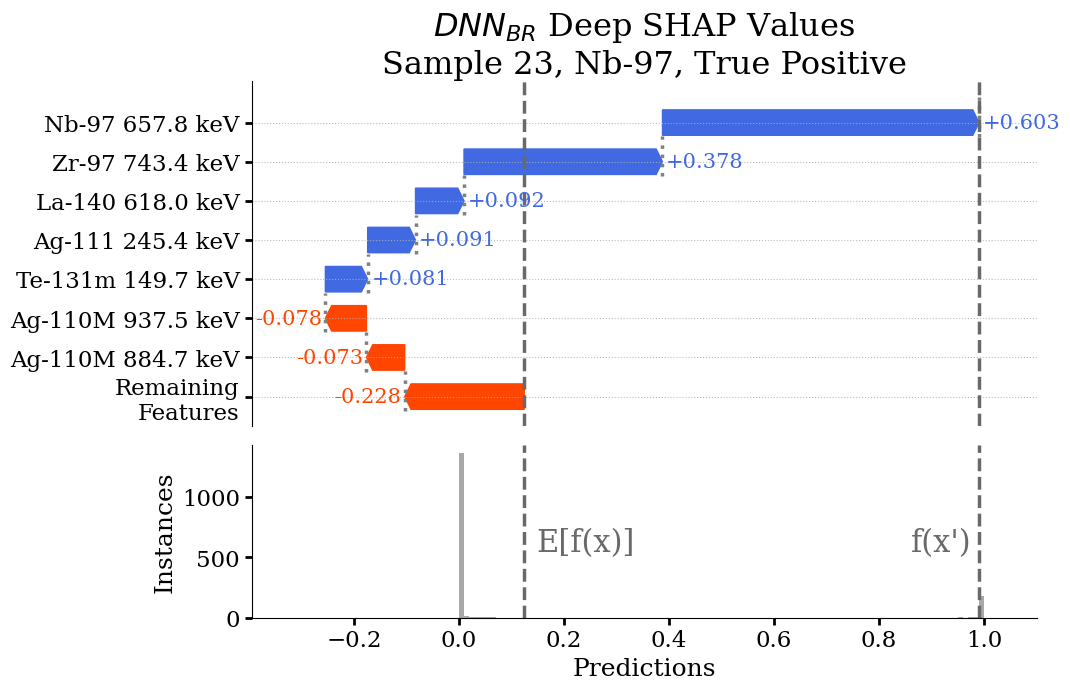"}
	\caption{Deep SHAP values for Nb-97 prediction made by a DNN$_{\text{BR}}$.}
	\label{fig:Nb97shap}
\end{figure}

For short-lived Nb-97, the primary photopeak of the parent isotope, Zr-97, was 
important in each identification. This aligned with our expectations because Nb-97 
not often found on its own (half life $\sim 72$ min) but rather as 
a decay product of Zr-97. The fact that this
correlation is captured by the ML models results in them producing fewer false positives 
than the methods of Genie 2000. A SHAP explanation of the Nb-97 prediction for the 
spectrum of Fig.~\ref{fig:Nb97_spec} shows the high importance of not only the Nb-97 peak but
also the Zr-97 peak (Fig.~\ref{fig:Nb97shap}). If the 743~keV peak of Zr-97 is not 
captured in a peak search (e.g., it might be buried by a nearby peak and not deconvolved
from it), this is detrimental to the Nb-97 identification and can
lead to a false negative, as observed in one instance. 

\begin{figure}[h]
	\centering
	\includegraphics[width=9cm]{"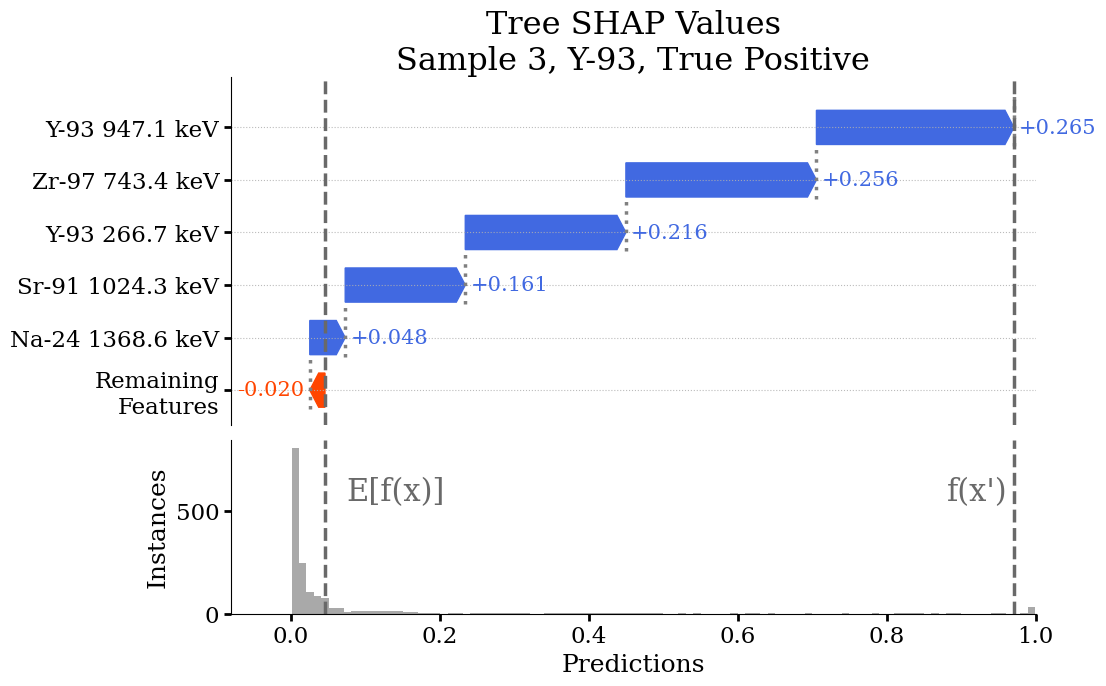"}
	\caption{Tree SHAP values for an XGB Y-93 prediction.}
	\label{fig:Y93shap1}
\end{figure}

\begin{figure}[h]
	\centering
	\includegraphics[width=9cm]{"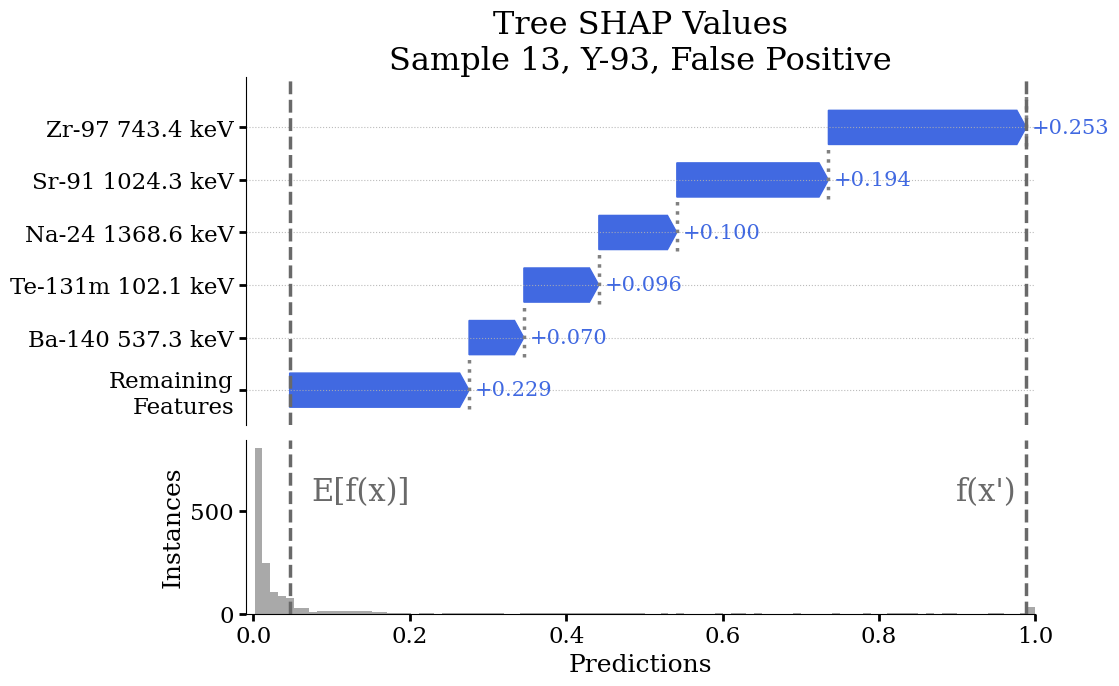"}
	\caption{SHAP values for an XGB false positive prediction. }
	\label{fig:Y93shap2}
\end{figure}

\subsection{Limitations}
\label{sec:limitations}
While the ML models in this work proved adept at NID,
their success was hampered when support for a given radionuclide was not strong. As seen
in the SHAP plot of Fig.~\ref{fig:Y93shap1} for Y-93, for example, the predictions strongly
depended on the presence of other short-lived isotopes. When Y-93 was not
detected in one case, the ML models still predicted 
its presence due to the more easily detected (i.e., longer lived and/or having stronger
emission intensities) isotopes such as Sr-91, Zr-97, and Na-24. This is shown 
in Fig.~\ref{fig:Y93shap2}.

In an effort to overcome this, we tested oversampling of classes with very low support 
to improve model performance metrics but found that this led to models less able to 
generalize to new gamma spectra containing one or more of those
isotopes due to the varied nature of those spectra. We believe  
adding high-fidelity synthetic spectra with varied content to the training data for 
classes with low support will be the best way to help models generalize.

We also found, unsurprisingly, that ML models became confused when data was 
inconsistently labeled. During this work, labeling mistakes were uncovered and 
amended precisely because of this fact. On the other hand, models did not seem to 
experience difficulty learning when photopeaks were present in the data that arose from
isotopes not included in the set of 65 nuclides for NID.

SHAP also has its own unique limitations. Primarily, SHAP is best used to assess the 
feature importances of a single example and maintains no regard for the ground truth label 
of a sample; it is is only concerned with how the prediction for one sample differs from 
the expectation of the overall dataset.

\section{Summary}
\label{sec:Sum}

In this work, we found that 
XGB and DNN models excel at mapping peak information from HPGe gamma spectra to 
NID predictions. 
Further, we've illustrated that SHAP values clearly explain the importance of specific 
input features in NID predictions. We have shown that the implemented ML 
classifiers make use of the most physically relevant peaks associated with specific 
isotopes for predictions and also rely upon the broader spectral context.

To illustrate model prediction explanations, 
we used SHAP to calculate feature importance. Tree and Deep SHAP values showed that:
\begin{enumerate}
	\item The strongest photopeak for any given isotope consistently
	had the largest SHAP value regardless of the ML model making a prediction,
	\item For nuclides with low support in the data set, e.g., Y-93, the most 
	important features for the predictions were from nuclides with similar 
	half-lives ($\sim 10$~hr for Y-93) and context,
	\item Models can leverage spectral context in
	predictions (e.g., peak from Zr-97 parent was important in predictions for Nb-97),
	\item XGB models value fewer input features more strongly in predictions than DNNs, 
	whose SHAP values are more evenly distributed.  
\end{enumerate}

Further, XGB and DNN models (of which XGB was best) both out-performed the NID method
with interference correction implemented in Genie 2000. This assessment was made 
using identical nuclear data (isotopes and peak energies) for the software-based NID 
to that leveraged in the model input space.
The software, using the methods of Refs.~\cite{Gunnink1972} and~\cite{Koskelo1981}, 
overestimated the presence of nuclides in HPGe gamma spectra. In contrast, both DNNs 
and XGB classifiers mitigated the number of false positives produced. 
Model generalizability was tested on recently collected NF R\&D spectra,
for which models achieved overall F1 scores of between 0.88 and 0.92, while 
the software achieved an F1 score of 0.80. 

The use of XGB and DNN models can ease the burden on spectroscopists in the selection of 
radionuclides. This is because  
these algorithms leverage the peak information to get a smaller, yet equally comprehensive
(i.e., nearly perfect recall) NID list when compared to the NID output of Genie 2000.

Avenues for expanding this research involve improving other parts of the spectroscopist 
work cycle or improving upon the models herein. Regarding the former, one could 
investigate ML-based ways to do peak finding and fitting that surpass traditional methods.
Further, the outputs of the models herein could be upgraded to reflect a  multi-label 
proportion prediction approach where peak areas are mapped to relative activities 
of the target nuclides. Regarding the latter, we believe this work could be rendered 
generally applicable by expanding
the dataset with high-fidelity simulations. Whereas in ML works using
complete spectra the discrepancy between simulated and experimental data is a nuisance,
peak areas extracted from high-fidelity simulations should
be near enough to their experimental counterparts to make that data very useful to ML
models and prepare them for a wide range of spectral contexts. Such an approach
could widen the gap by which ML models already outperform contemporary NID
methods.

\section*{CRediT Authorship Contribution Statement}
\textbf{Samuel Emmons:} Conceptualization, Methodology, Formal analysis, Investigation,
Data Curation, Writing - Original Draft, Writing - Review \& Editing, 
Visualization, Supervision
\textbf{Kelly Truax:} Methodology, Validation, Formal analysis, Investigation,
Data Curation, Writing - Original Draft, Writing - Review \& Editing, Visualization
\textbf{Maurice Lonsway:} Methodology, Validation, Formal analysis, Investigation,
Data Curation, Writing - Original Draft, Writing - Review \& Editing, Visualization
\textbf{Bruce Pierson:} Conceptualization, Supervision, Funding acquisition
\textbf{Brian Archambault:} Conceptualization, Resources, Writing - Review \& Editing,
Supervision, Project administration, Funding acquisition

\section*{Acknowledgements}
This work was supported by the Pacific Northwest National Laboratory (PNNL) Laboratory 
Directed Research and Development Program and is a contribution of the Nuclear 
Forensics Transformational Innovation (NFTI) Initiative. A portion of the research 
was performed using resources available through Research Computing at PNNL. PNNL is 
a multiprogram national laboratory operated by Battelle for the Department of Energy 
under Contract No.~DE--AC05--76RLO~1830. 
We also thank the Radiological Processing Laboratory (RPL) service center,
as well as the count room staff in the 3420 Radiation Detection Laboratory, at PNNL for 
gamma emission analysis data. We also thank the teams of PNNL chemists 
involved in sample preparation. Finally, we thank Matt Douglas and Lori Metz for their
leadership of the NFTI Initiative.

\appendix
\setcounter{equation}{0}
\setcounter{figure}{0}
\section{Additional ML Explanations}
\label{sec:AppendixA}
\begin{figure}[h]
	\centering
	\includegraphics[width=9cm]{"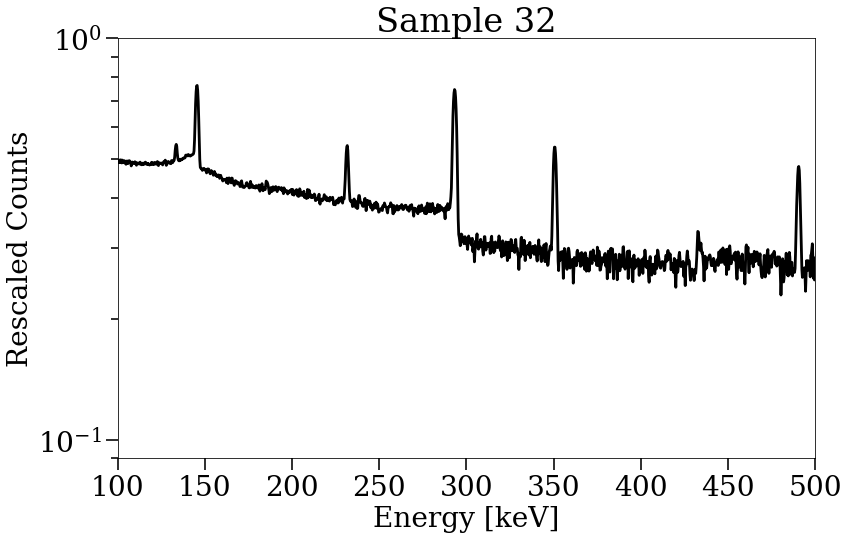"}
	\caption{Spectrum containing isotopes of Ce and other elements.}
	\label{fig:ce_spec1}
\end{figure}
\begin{figure}[h]
	\centering
	\includegraphics[width=9cm]{"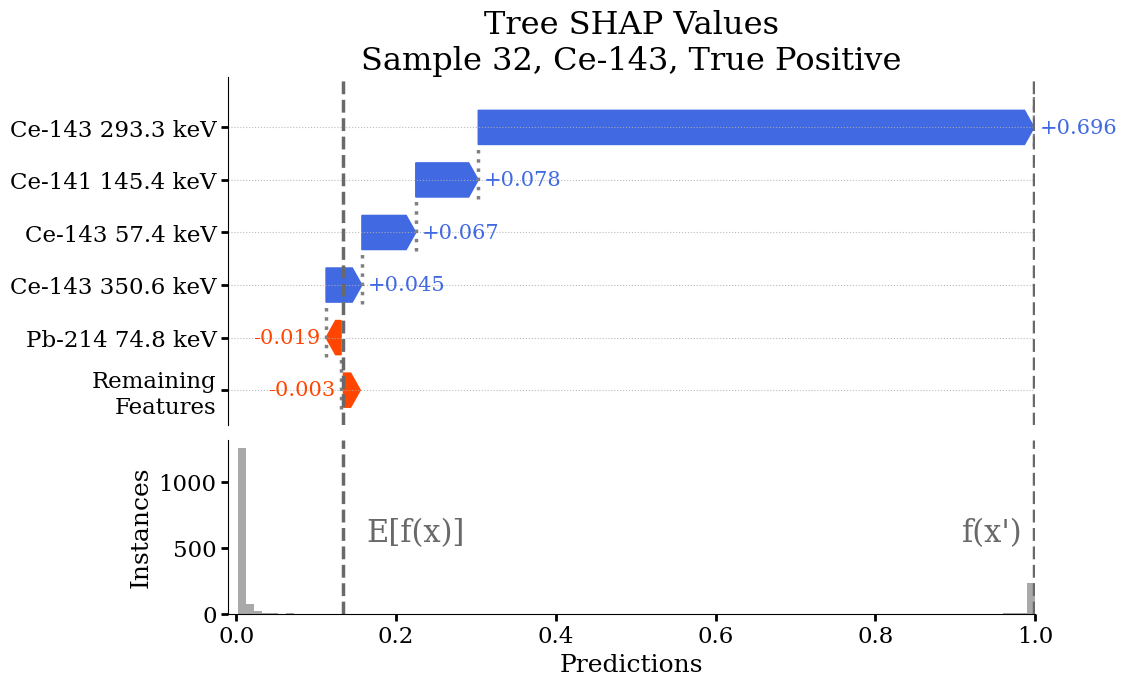"}
	\caption{Tree SHAP explanation of $^{143}$Ce XGB prediction in sample that predominantly
		contained Ce isotopes. }
	\label{fig:ce143_tp_32}
\end{figure}

Here, we provide SHAP plots for another case study: the fission product isotopes 
of Ce (Ce-141, 143, and 144). In complex samples produced 
in NF R\&D experiments, these isotopes are often, but not always, simultaneously observed. 
For example, short-lived fission products can cover up the two longer-lived of the Ce 
radioisotopes (particularly Ce-144). In other cases, these isotopes 
occur in samples containing a reduced set of elements  
in which they are each readily observed. Importantly, we found that the 
ML models trained in this work easily distinguish the three isotopes from each other.

\begin{figure}[t]
	\centering
	\begin{subfigure}[b]{0.48\textwidth}
	\includegraphics[width=\textwidth]{"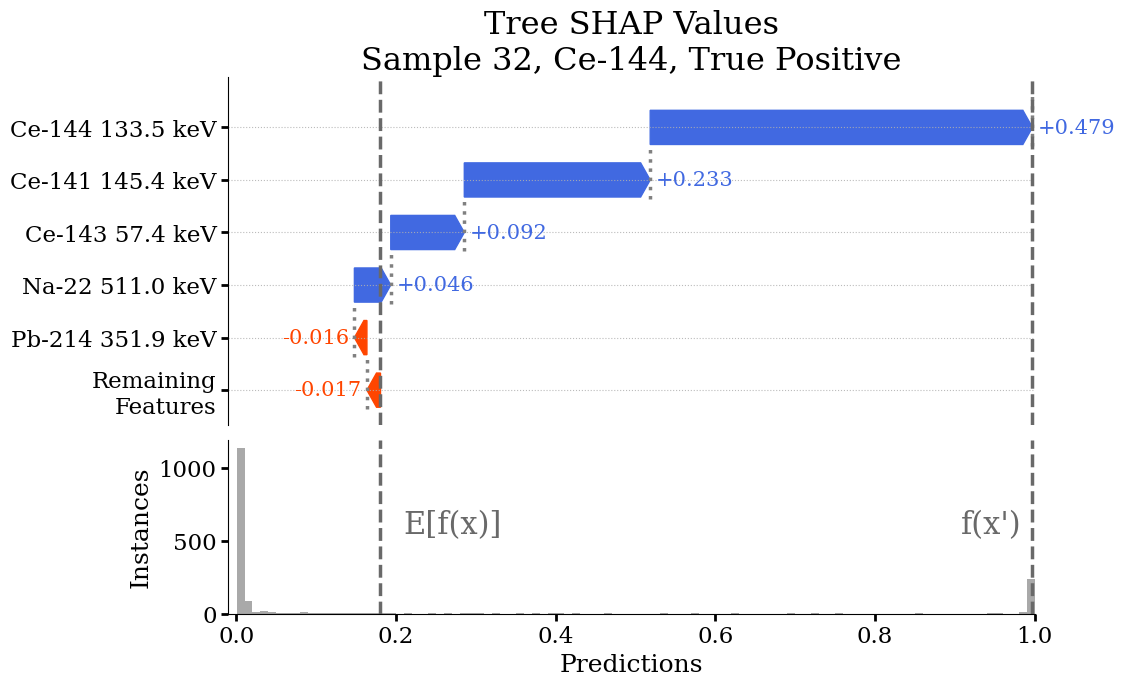"}
	\end{subfigure}
	\hfill
	\begin{subfigure}[b]{0.48\textwidth}
	\centering
	\includegraphics[width=\textwidth]{"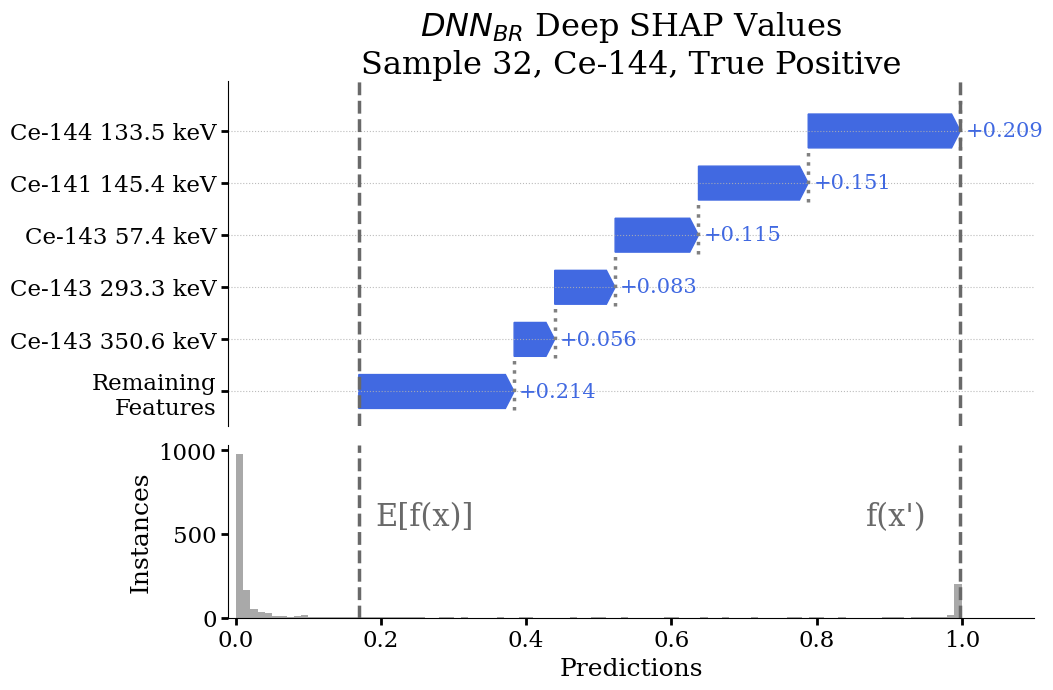"}
	\end{subfigure}
	\caption{Ce-144 predictions explained  
		for two different ML models (XGB and DNN$_{\text{BR}}$) by their respective SHAP methods.}
	\label{fig:ce144_tps}
\end{figure}

\begin{figure}[t]
	\centering
	\includegraphics[width=9cm]{"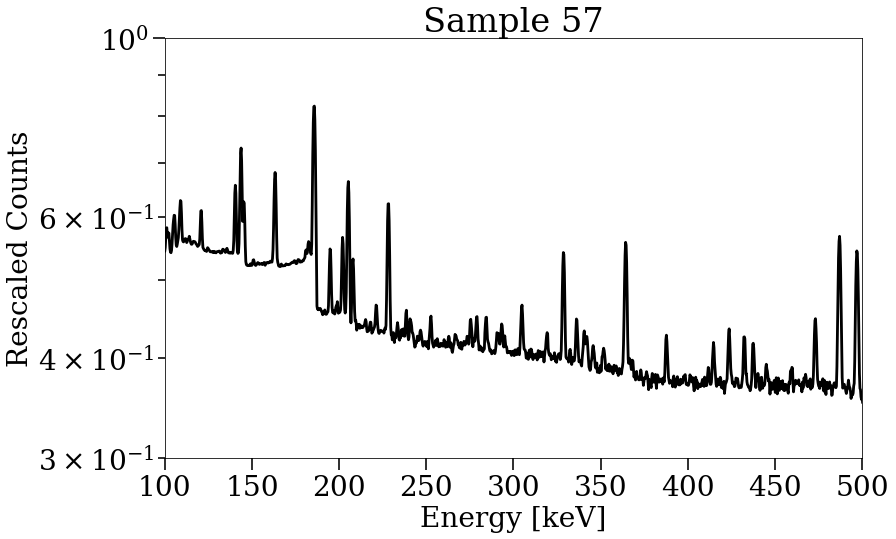"}
	\caption{Spectrum containing isotopes of Ce and other elements.}
	\label{fig:ce_spec2}
\end{figure}

The spectrum of Fig.~\ref{fig:ce_spec1} primarily contains Ce isotopes.
Each of the ML models produced true-positive predictions for the three Ce isotopes. 
\begin{figure}[h]
	\centering
	\includegraphics[width=9cm]{"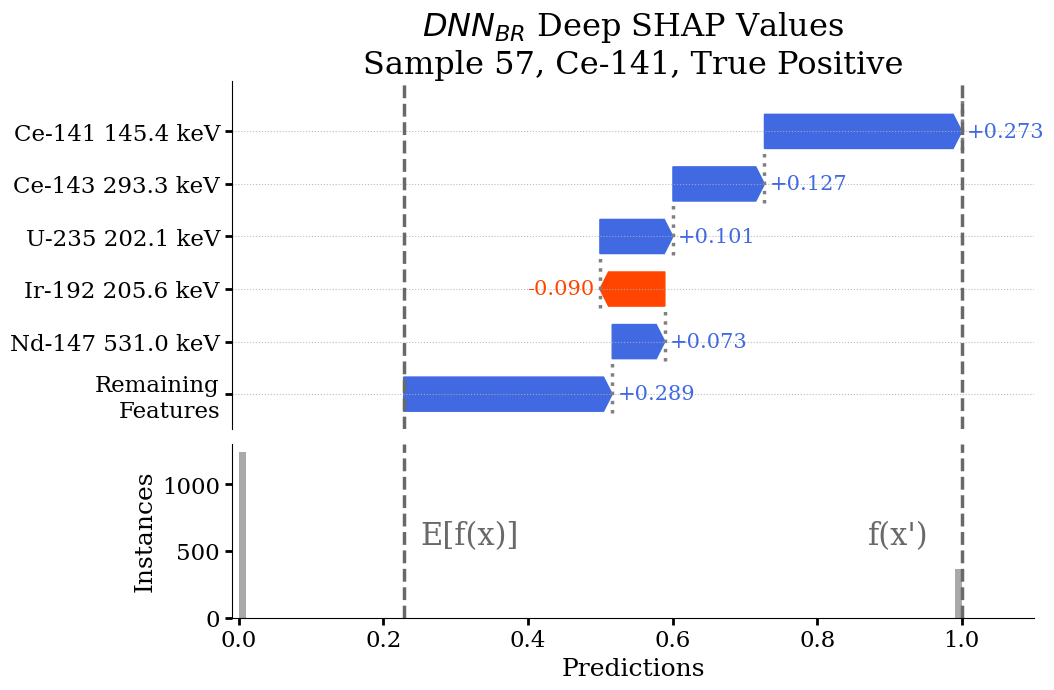"}
	\caption{Deep SHAP explanation of feature importance for $^{141}$Ce prediction.}
	\label{fig:ce141_2_tp}
\end{figure}
Figure~\ref{fig:ce143_tp_32} and the left panel of Fig.~\ref{fig:ce144_tps} 
show Tree SHAP values explaining 
XGB predictions for Ce-143 and 144, respectively. In the first case, the primary 
peak associated with Ce-143 was by far the predominant feature for classification. 
For Ce-144, which is infrequently observed by itself in our training dataset,
the peak associated with Ce-141 was also strongly relied upon. Though this may seem like a 
drawback at first, it is an important part of the context in our
data for determining whether or not Ce-144 is present.

The other models also leveraged the spectral context when making predictions for 
the isotopes of Ce. The Deep SHAP explanation in the right panel of~Fig.~\ref{fig:ce144_tps} 
shows that the peaks for Ce isotopes \textit{besides} Ce-144 were proportionally 
more important to the DNN$_\text{BR}$ than they were to the XGB model. 
Feature dependence relationships similar to 
those of the DNN$_{\text{BR}}$ were also observed in the multi-label DNN. This was expected
because of the way XGB and DNN models learn to operate on the input space. The decision
trees of XGB learn to ignore irrelevant features in a given 
classification decision, whereas information from all input nodes is passed along in a DNN.

ML models were also able to identify the isotopes of Ce in HPGe gamma 
spectra with many additional photopeaks. The spectrum shown in Fig.~\ref{fig:ce_spec2}
is an example of such a case. Here, the photopeaks of the Ce isotopes are not large in
comparison to other peaks in the spectrum. The primary peak associated with Ce-143 is
particularly small. We found that in a few situations like this, the models 
occasionally produced a prediction probability that was close to, but below, the prediction 
threshold used (i.e., a false negative occurred). Additional support in the training
data that includes the isotopes in a wider range of ratios would likely mitigate this
issue.

\begin{figure}[h]
	\centering
	\begin{subfigure}[b]{0.48\textwidth}
	\includegraphics[width=\textwidth]{"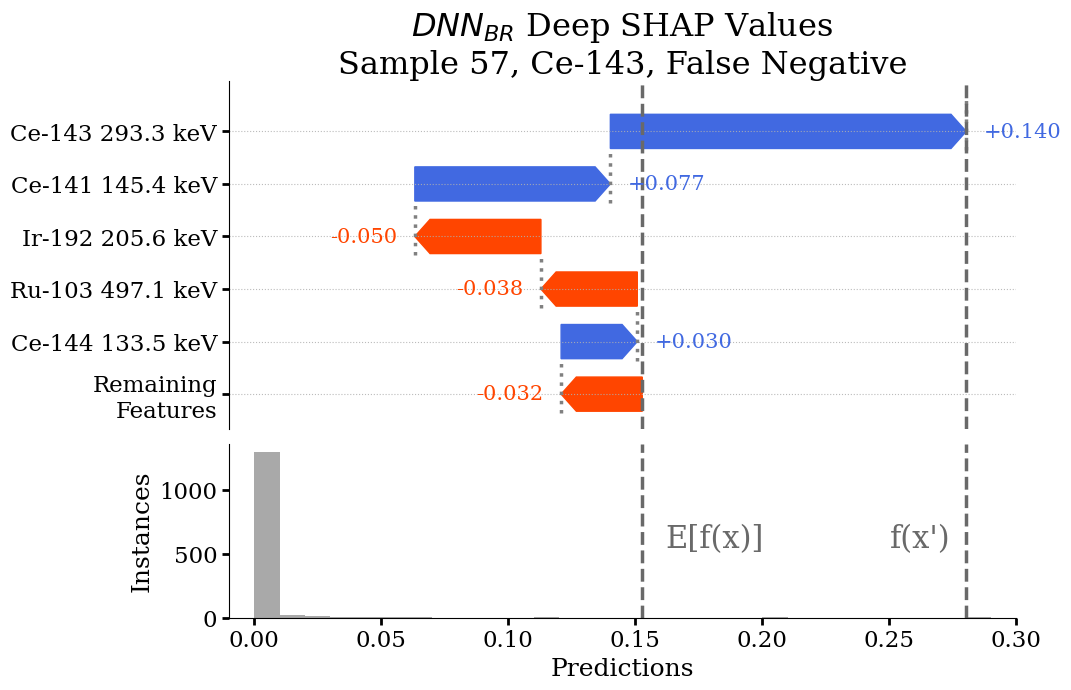"}
	\end{subfigure}
	\hfill
	\begin{subfigure}[b]{0.48\textwidth}
	\centering
	\includegraphics[width=\textwidth]{"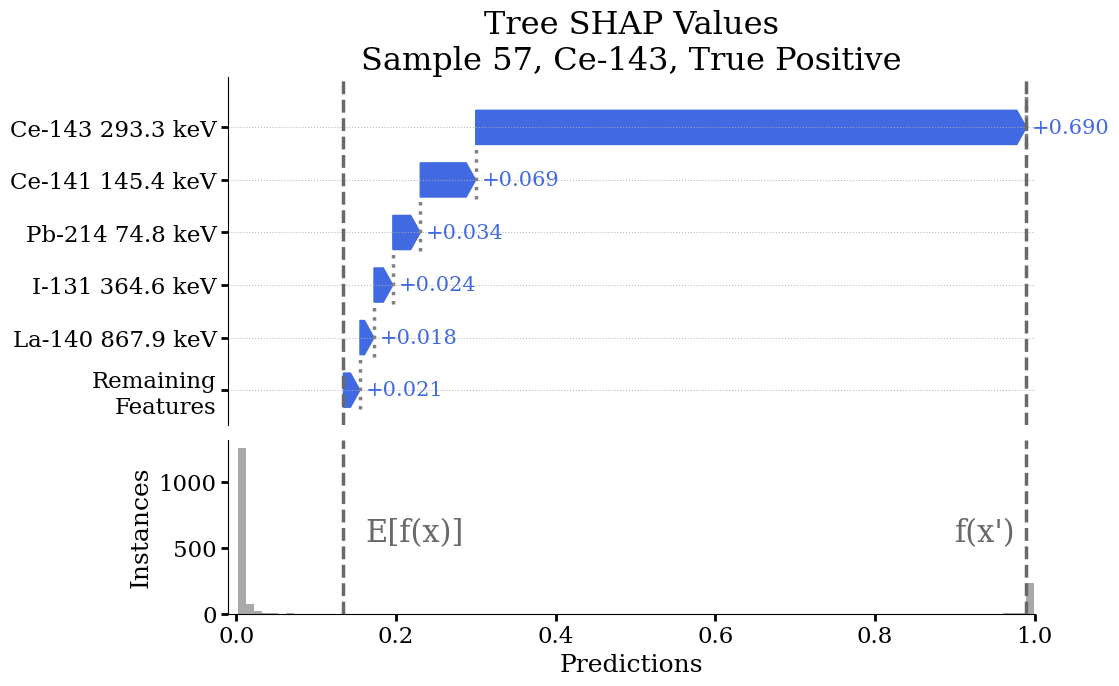"}
	\end{subfigure}
	\caption{SHAP explanations of DNN and XGB model predictions of Ce-143 in the spectrum
	shown in Fig.~\ref{fig:ce_spec2}.}
	\label{fig:ce143_tps}
\end{figure}
Deep SHAP plots explaining the DNN$_{\text{BR}}$ predictions for Ce-141 and 143 in 
the spectrum of Fig.~\ref{fig:ce_spec2} are shown in Figs.~\ref{fig:ce141_2_tp} 
and the left panel of Fig.~\ref{fig:ce143_tps}. 
For each of these plots, the correct photopeak has the primary
importance. In the case of the second plot, the DNN$_{\text{BR}}$ was close to the right
answer, but the Ce-143 feature was small enough such that it did not drive the 
prediction past the prediction threshold of 0.4. 
On the other hand, the XGB model easily made the correct prediction, as 
shown in the right panel of Fig.~\ref{fig:ce143_tps}, by leveraging the primary peak for Ce-143. 
Notably, the \textit{second most} important feature in each
prediction was from another Ce isotope for each of the predictions shown, 
illustrating that other isotopes of Ce were more contextually important for the 
predictions than other co-occurring fission products.

\FloatBarrier

\bibliographystyle{elsarticle-num}

\end{document}